# The Fluctuations of Blocked Ionic Current Reveal the Instantaneous Statuses of DNA in Graphene Nanopore

*Wenping Lv, Ren'an Wu*[*]

CAS Key Lab of Separation Sciences for Analytical Chemistry, National Chromatographic R&A Center, Dalian Institute of Chemical Physics, Chinese Academy of Sciences (CAS), Dalian, 116023, China

**The corresponding Authors Footnotes:**

Ren'an Wu (wurenan@dicp.ac.cn)

Tel: +86-411-84379828; Fax: +86-411-84379617

Wenping Lv (wenping@dicp.ac.cn)

Tel: +86-411-84379617; Fax: +86-411-84379617

## Abstract:

Extracting the sequence information of DNA from the blocked ionic current is the crucial step of the ionic current-based nanopore sequencing approaches. The thinnest graphene nanopore, which contained only one layer of carbon atoms, potentially has ultra-high DNA sequencing sensitivity. However, the dynamical translocation information of DNA contained in the blocked ionic current has not been well understood to date. In this letter, an assessment to the sensitivity of ionic current-based graphene nanopore DNA sensing approach was carried out using molecular dynamics simulations. By filtering the molecular thermal motion induced noise of ionic current, we found that the instantaneous conformational variations of DNA in graphene nanopore could be revealed from the fluctuations of the denoised ionic current. However, the blockage of ionic current which induced by the proximity of the DNA base-pairs to the nanopore (within 1.5 nm) was also observed. Although the expected single-base resolution of graphene nanopore should be enhanced by further studies, our findings indicated that the ionic current-based graphene nanopore sensing approach has high sensitivity to the instantaneous translocation status of DNA.

Nanopore sequencing is a new technology promising to directly read out the gene information of DNA at single-molecule level.[1, 2] The center stage of nanopore sequencing is to distinguish the signals of different kinds of bases of DNA through a nanopore.[3-9] The subnanometer thickness (0.34 nm) of graphene sheet comparable to the spatial interval of DNA nucleotide suggests that the nanopore sequencing at single-base level could be realized utilizing a graphene nanopore.[4, 10] Based on graphene nanopores created experimentally,[11] the translocation of double-stranded DNA (dsDNA) through monolayer and/or multilayer graphene nanopores has been recently demonstrated.[5-7] In these experiments, the fluctuation of blocked ionic current was observed, explained as the difference induced by folded/unfolded DNA or the unzipping of DNA chains.[5-7] Thanks to the atomic level molecular dynamics (MD) simulation technology, the subtle structural features of DNA and the graphene-DNA interactions during translocation could be further revealed. In 2011, Schulten et al. observed the difference of ionic current blockages which induced by the folded/unfolded DNA, and suggested that under suitable bias conditions A-T and G-C base-pairs can be discriminated using graphene nanopores.[12] Recently, Aksimentiev et al. reported that the translocation of single-stranded DNA (ssDNA) through graphene nanopores might occur in single nucleotide steps,[13] similar with the biologic nanopores. However, the sensitivity of ionic current blockades to the orientations of nucleotides in graphene nanopore has also been observed.[13, 14]

Actually, ionic current-based nanopore sensing relies on ions through a nanopore which contribute to both the signal and the noise.[15-19] In particular, the membrane capacitance produces noise fluctuations that increase with the bandwidths of measurement,[3] and the noise of ionic current in membrane-like graphene nanopore was distinctly huger than that in the channel-like synthetic nanopore.[12, 20] To reduce the electrical noise of graphene nanopore, a stacked graphene-$Al_2O_3$ nanopore was constructed and the temporal resolution of DNA and/or DNA-protein complexes detection was significantly improved recently.[21] Based on the differences of ionic current, researchers found that the translocation of DNA in nanopores usually accompanied with the deformation of DNA.[12, 20, 22-24] Due to the atomic thickness of graphene, the graphene nanopore sensors might have ultra-high sensitivity to the instantaneous translocation statues of DNA. However, the translocation information of DNA contained in the fluctuation of blocked ionic current has not been well understood to date.

Therefore, a systematic MD simulation study was presented in this letter, to explore the sensitivity of ionic current to the instantaneous translocation statuses of DNA within graphene nanopore. Before we extract the translocation information of DNA from ionic current, the ionic current measurement itself ($I(t) = \{\sum_{i=1}^{N} q_i[z_i(t + \Delta t) - z_i(t)]\}/\Delta tLz$) was assessed by investigating the fluctuation (root-mean-square, RMS) dependencies of ionic current to the measure interval ($\Delta t$)[16, 20], simulation temperature and bias voltage. By monitoring the variations of local conformation, in-pore translocation velocity and graphene-DNA interaction of DNA within graphene nanopore, how the conformational, dynamical and interactional information of in-pore DNA revealed from the fluctuation of ionic current signal were presented. We found that 1) the synchronous change of the number of atoms of DNA accumulated in graphene nanopore and the blockage of ionic current was directly showed after the thermal noise of ionic current has been filtered; 2) the blockage of ionic current could also be induced by the proximity of DNA base-pairs to the nanopore. To the best of our knowledge, it should be the first reported result dynamically shows the high sensitivity of graphene nanopore to the instantaneous translocation statuses of in-pore DNA.

As shown in Figure 1a, the interactions among ions, water molecules and graphene directly impact the motion of the charge carriers (ions) in NaCl solution.[25] Therefore the impact of measurement frequency ($1/\Delta t$) to the fluctuation of open-pore ionic current was investigated with different temperatures (280K, 300K and 320K), bias voltages (0V and 1V), graphene models (flexible and rigid) and cell dimensions (10 nm and 20 nm in z-direction). The obtained average (AVG) and fluctuation (root-mean-square, RMS) of ionic currents were plotted as function of the measure frequency ($1/\Delta t$) in Figure 1b-d. If no external bias voltage was applied (0V, 300K), the average ionic current was maintained in zero because no directional movement of ions was occurred in the system. While the fluctuation of ionic current was increased with the rise of measure frequency linearly, indicating that the molecular thermal motion (no bias voltage) induced Johnson-Nyquist (thermal) noise of ionic current was sensitive to the choice of measure frequency. After the bias voltage (1V, 300K) was applied, the fluctuation of the ionic current was even higher than the average ionic current (8 nA) when $\Delta t$ was shorter than 4 ps (250 GHz), and it was undistinguishable with net thermal noise (0V, 300K). With the decrease of measure frequency, the fluctuation of ionic current was reduced, but it still obviously greater than net thermal noise, suggesting that the bias voltage could also induce the enhancement of the noise of ionic current.[15] Comparing the results of different simulation temperatures (280K, 300K and 320K), the temperature sensitivity of both average and fluctuation of ionic current were distinctly presented (Figure 1b), suggesting that in a certain bias voltage the temperature determined molecular thermal motion contributes significantly to the ionic current. Although the ions would also accumulate on the surface of graphene nanopore for the oppression of applied electric field,[12, 25] but the impact of the shaking of carbon atoms at graphene nanopore edge (Figure S2) to the fluctuation of ionic current (Figure 1c) was not as obvious as the influence of temperature (Figure 1b). By the way, similar with the reported study for biologic nanopore (α-Hemolysin),[26] the average ionic current could maintain steady only if the time interval of measurements were shorter than 50 ps (> 20 GHz). The abnormal drop of average ionic current might result from the periodic boundary condition (PBC) employed in MD simulations, because it has been effectively alleviated (Figure 1d) by using a bigger simulation cell (20 nm in z-direction). These findings suggest that the “signal-to-noise” ratio could be improved by modulating the measure frequency, analogous to the experimental and theoretical results that the thermal noise of ionic current increases with the bandwidth of a detector.[16, 18, 24-26]

Based on above discussions, the measure interval of ionic current in the following studies was chose as 50 ps to ensure the “signal-to-noise” ratio > 5 (Figure 1d). The microscopic kinetics of a dsDNA chain (d-poly(CAGT)$_{48}$) electrophoretically passing through a 2.4 nm monolayer graphene nanopore were investigated based on 6 sections of MD simulations (indexes 10-15 in Table S1). As shown in Figure 2a, the original ionic current signals (grey lines) were further denoised with a FFT filter (cutoff frequency was 10GHz) to remove the impact of the thermal noise (red lines). Therefore, the presentation ability of ionic current to the instantaneous translocation statues of DNA in graphene nanopore was improved (blue lines). The profiles of the denoised ionic current (blue lines) were extremely different in the repeat MD simulations (R1, R2 and R3), suggesting that the translation statuses of DNA in graphene nanopore might be different in these simulations. To capture the instantaneous dynamical information of DNA in graphene nanopore, the in-pore translocation velocity of DNA was monitored:

$$\mathrm{V}(t) = \frac{1}{N(t)\Delta t} \sum_{i=1}^{N(t)} [z_i(t + \Delta t) - z_i(t)] \quad (1)$$

Namely, the in-pore velocity of DNA was defined as that within time interval of Δt the displacement of the part of DNA located in graphene nanopore with a length of ΔZ; $N$(t) represents the number of atoms of in-pore DNA at time point of $t$. The ΔZ was chose as 1 nm in calculation. As shown in panels of [1V, R1], [1V, R2], [2V, R1] and [2V, R2] in Figure 2b, although the bias voltage used in the calculations were 10 times higher than that in experiments,[1, 5, 6, 12] the translocation velocities of in-pore DNA were fluctuated around 2 mm/ms (5.5 kbp/ms) and around 3 mm/ms (8.5 kbp/ms) in most of the translocation time for 1 V and 2 V bias voltages, respectively. The translocation velocities of DNA in these simulations seem to be comparable with that of DNA which obtained experimentally using solid-state nanopores.[27-30] While there also were some unpredictable quick translocation events presented in these results, suggesting that the translocations of DNA in graphene nanopore were unstable. Especially, similar with the translocation of ssDNA in graphene nanopores,[13] the unceasing velocities fluctuations of DNA in simulations of [1V, R3] and [2V, R3] showed that the translocation of dsDNA could also be stagnated in graphene nanopore. The corresponding ionic current signals were fluctuated around 3 nA and 3.5 nA, respectively. While for the other four results in Figure 2a, the magnitude of ionic current signals were rose to the level of open-pore ionic current after about 12 ns ([1V,

R1]), 10 ns ([1V, R2]), 6 ns ([2V, R1]) and 5 ns ([2V, R2]), respectively. These results indicated that not only the initial conformations of DNA (folded/unfolded), the instantaneous translocation statues of DNA could also impact the blockaded ionic current significantly.

Therefore, the undulates of the denoised ionic current (Figure 2a) were further investigated to explore more detailed instantaneous translocation information of DNA. As an example, the peaks and troughs of the denoised ionic current of [2V, R1] in Figure 2a were marked with arrows a-g in Figure 3. Results show that the peaks and troughs of blocked ionic current were corresponding to different local conformations of DNA in graphene nanopore one by one (insets a-e of Figure 3). Meanwhile, the trajectory of MD simulation (Movie S1) also dynamically shows that the instantaneous conformational variations (such as yawing and upright) of in-pore DNA did accompanied with the fluctuations of blocked ionic current. Different with the stacking interaction-induced stepwise translocation of ssDNA in graphene nanopore,[13] the instantaneous conformational variations induced velocity fluctuations of dsDNA (Figure 2b) were more like a DNA deformation-induced translocation jam. The comparison between the number of atoms of DNA accumulated in graphene nanopore, N(t), violet region of DNA in Figure 4a, and the ionic current signals (Figure 4b and Figure S3) directly show that the fluctuations of blocked ionic current were reciprocal to N(t) elaborately for all the non-stagnant translocation events ([1V, R1], [1V, R2], [2V, R1], [2V, R2]). However, a plane parameter "effective unoccupied area", which was proposed in a recent publication, could only basically reveal the spatial blockage effect of DNA to the fluctuation of ionic current, because the spatial blockage effect of DNA towards ionic current was reduced into a 2-dimensional (2D) parameter in their model.[31] These results indicate that not only the occupied area of nanopore, the instantaneous conformational variations of the in-pore DNA was also a key factor of the fluctuation of ionic current blockages.

The unstable translocations of DNA which revealed from the ionic current suggest that the interactions between DNA and graphene were varied with the translocation. As shown in Figure 4c, the DNA-graphene interaction was fluctuated in range of -25

~ -150 kJ/mol in the first 3 ns. Compared with the average interaction energy (-27.08 ± 6.32 kJ/mol) between a short in-pore DNA fragment composed of only two base-pairs(A*p*T and G*p*C) and the same graphene nanopore (aperture 2.4 nm, monolayer) in previous research,[24] the enhanced DNA-graphene interaction imply that the nucleobases were exposed toward graphene surface (π-π stacking interaction was much greater than edge-edge interaction between nucleobases and graphene[32, 33]), and/or the neighbor DNA base-pairs contributes to the DNA-graphene interaction also. The MD trajectories (Movie S1-S2) showed that the exposed nucleobases did adhered on the bottom of graphene nanopore steadily after 3 ns, according with the dramatically enhancement of DNA-graphene interaction after 3 ns (inset of Figure 3c). By the way, due to the strong π-π stacking interaction, the in-pore translocation velocity of DNA was also reduced after 3ns, and the DNA translocation was almost stagnated around 4 ns ([2V, R1] in Figure 2b). While, the energy barrier of the bending of DNA and the strong electrostatic force applied on DNA which near to graphene nanopore[12, 24] induced the remaining DNA crosswise lying down and blocked graphene nanopore closely (insets e-f of Figure 3 and Movies S1-S2). Thus the adherence of DNA on graphene could also induce the unexpected fluctuations of ionic current (marked with arrows e-f in Figure 3). After DNA passed through the graphene nanopore (6~8 ns), the magnitude of ionic current (Figure 3) was rose to the level of open-pore ionic current (> 10 nA), and the structural fluctuation of graphene nanopore (highlighted with yellow band in Figure 4d) was also decreased to the level of no DNA system (Figure S2). Thus the fluctuations of the blocked ionic current might also be influenced by the DNA-graphene interaction induced structural fluctuations of graphene nanopore.

As we suggested above, the translocation statues of DNA base-pairs near to graphene nanopore could also influence the fluctuation of ionic current. Thus a DNA fragment composed of only two base-pairs (d-$(AG)_2$) was employed as a stopper to probe the blockage effect of the neighbor base-pairs of DNA at pore entrance. A set of MD simulations (indexes 16-59 in Table S1) were performed to get the ionic blockage effect of the stopper at different positions and orientations. The shift range of the stopper (probe DNA) to the center of graphene nanopore was -2 nm ~ 2 nm (see insets of Figure 5a). The schematic diagrams of the orientation altering of probe DNA were shown as insets of Figure 5b. The obtained ionic blockages which induced by the probe DNA with two orientations were shown in Figure 5a. We found that the blocked ionic current ($I$) induced by DNA within graphene nanopore (distance was 0 nm) was only about half of the open-pore current ($I_o$), it accords with the reported result.[14] The interesting result was that the DNA near to graphene nanopore could also induce the ionic current blockages. For instance, when the probe DNA was positioned within 0.4 nm to graphene nanopore in z-direction, the blocked ionic currents were almost equal to a half of the open-pore current ($I_o$), suggesting that the blockage effect of DNA which near to the entrance of graphene nanopore towards ionic current ($I/I_o$) was similar to that of DNA within nanopore (Figure 5b). Additionally, the ionic current blockages were enhanced with the decrease of the DNA-graphene interval (Figure 5a) when the probe DNA was positioned within 1.5 nm of graphene nanopore. The neighborhood effect of DNA to ionic current blockage indicated that the conformational variation of the neighbor base-pairs of DNA around graphene nanopore entrance could also induce the fluctuation of ionic current.

In summary, a series of MD simulations were carried out to assess the ionic current measurement and to extract the instantaneous translocation information of DNA in graphene nanopore from the fluctuations of blocked ionic current. A key result of our study was that the instantaneous conformational variations of in-pore DNA were synchronously revealed from the undulations of denoised ionic current because the fluctuation of the number of atoms of DNA accumulated in graphene nanopore. However, we also found that both the DNA base-pairs within and near to the entrance of graphene nanopore have similar blockage effect to ionic current. Compared with other sensing approaches based on graphene (*i.e.* the transverse conductance measurement of graphene nanopore,[34] nanoelectrode[35] and nanoribbon[36] etc.), the DNA base-specific resolution of the ionic current-based graphene nanopore sensing should be further improved. Modifying the graphene nanopore with functionalized groups,[35] might be a potential strategy to enhance the DNA base distinguish ability of ionic current based on graphene nanopore sequencing system.

## Acknowledgement:

This work was supported by the National Natural Science Foundation of China (No. 21175134), the Knowledge Innovation Program of Dalian Institute of Chemical Physics and the Hundred Talent Program of the Chinese Academy of Sciences to Dr. R. Wu.

## Supporting information Available:

Detailed description of the simulation methods; plot of the average and fluctuation of the ionic current obtained from the three benchmark MD simulations by using different temperature coupling methods (nose-hoover, berendsen and v-rescale); plot of the root-mean-square distances (RMSD) of the flexible and rigid graphene nanopores; plots of the comparing of ionic current and number of DNA atoms in graphene nanopore for the simulation [1V, R1], [1V, R2], [1V, R3], [2V, R2] and [2V, R3]; list of the detail simulation parameters of all the calculations in our work; animations illustrating of the synchronically evolution of the MD trajectory of d-poly(CAGT)$_{48}$ DNA translocation in graphene nanopore and the fluctuation of blocked ionic current as well as the molecular details of DNA adhering on graphene surface.

# Figures and Legends:

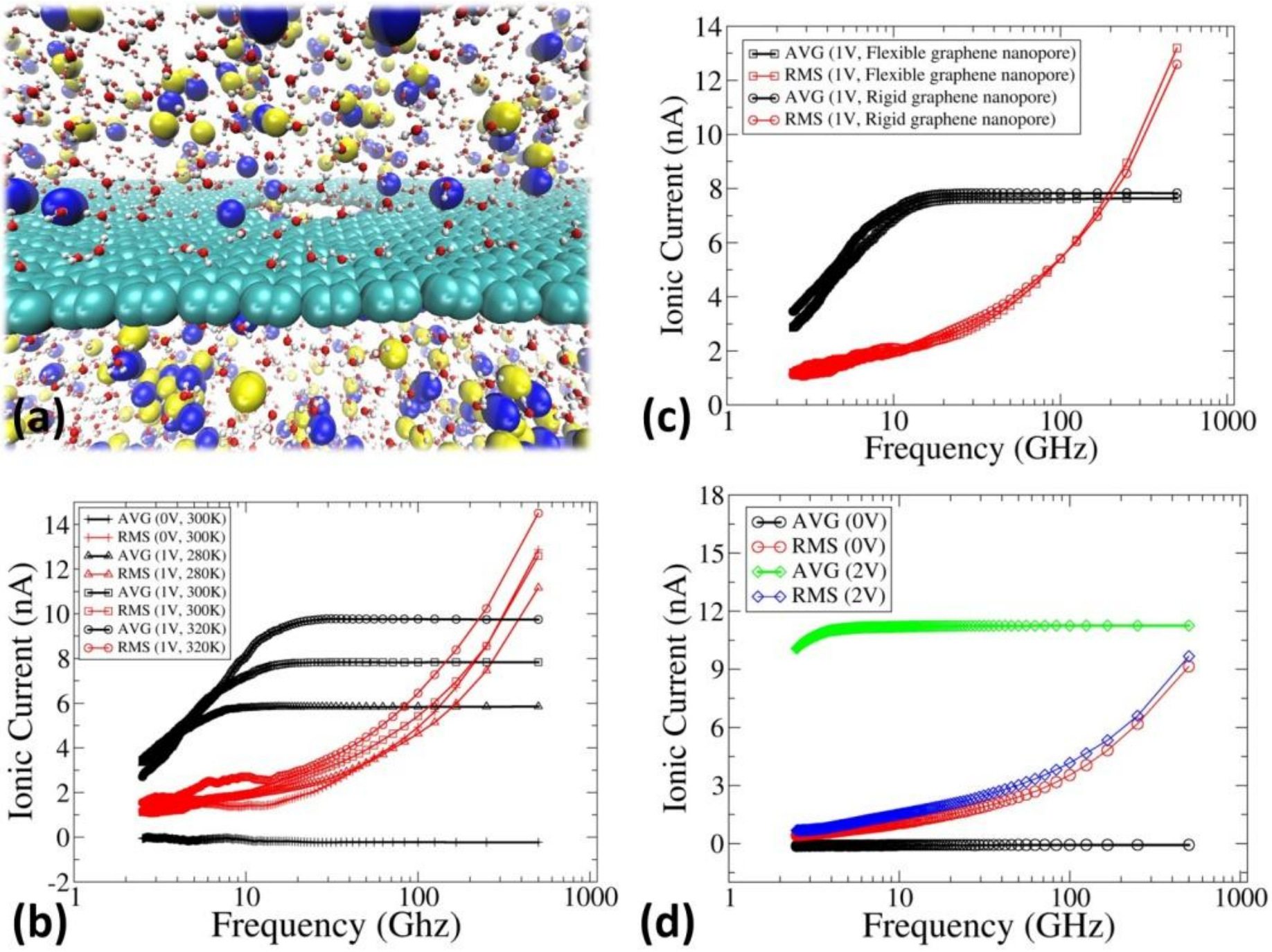

Figure 1. (a) The schematic diagram of open-pore ionic current simulation. The $Na^{+}$, $Cl^{-}$ and graphene nanopore were colored with yellow, blue and cyan in "VDW model". For representation convenience, here only 10% of the water molecules in simulation system were showed in "CPK model". (b-d) The average (AVG) and the fluctuation (root-mean-square, RMS) of open-pore ionic current for the simulations of (b) different temperature and bias voltage, of (c) flexible and rigid graphene nanopores and of (d) the big simulation box (6.3x6.3x20 $nm^3$) were plotted as function of the measure frequency ($1/\Delta t$). Data were obtained from the last 3-ns of the 4-ns MD trajectories.

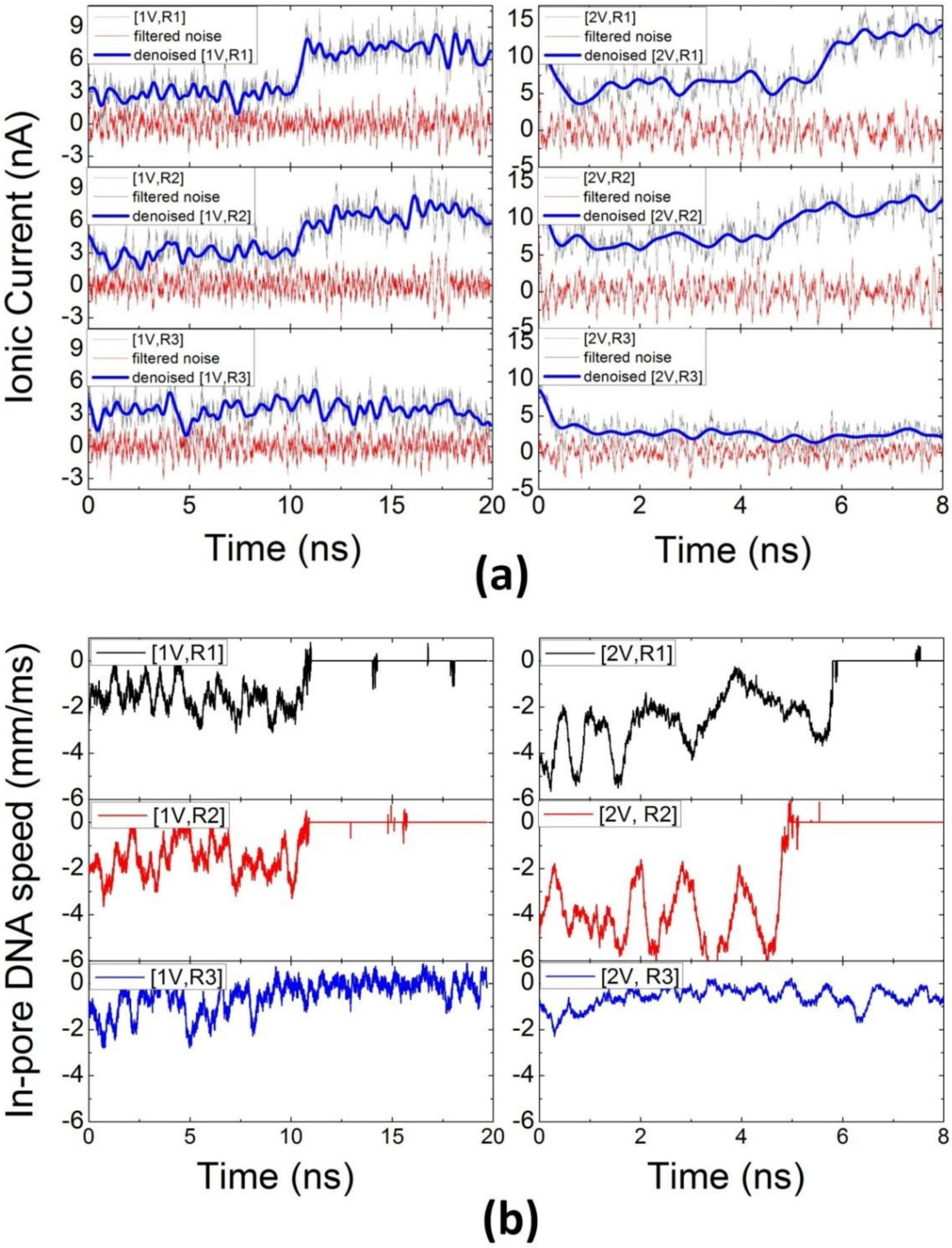

Figure 2. (a) The ionic current signals (grey lines) which obtained from the MD simulations of 10-15 in Table S1 were plotted as function of simulation time. The denoised ionic current signals (blue lines) were smoothed by a FFT filter with cutoff frequency of 10GHz. The filtered noises were also presented (red lines). (b) The instantaneous translocation velocity of DNA in graphene nanopore was plotted as function of simulation time. In all the legends, the 1V and 2V were the applied bias voltages; the R1, R2 and R3 were used to label the three repeated MD simulations.

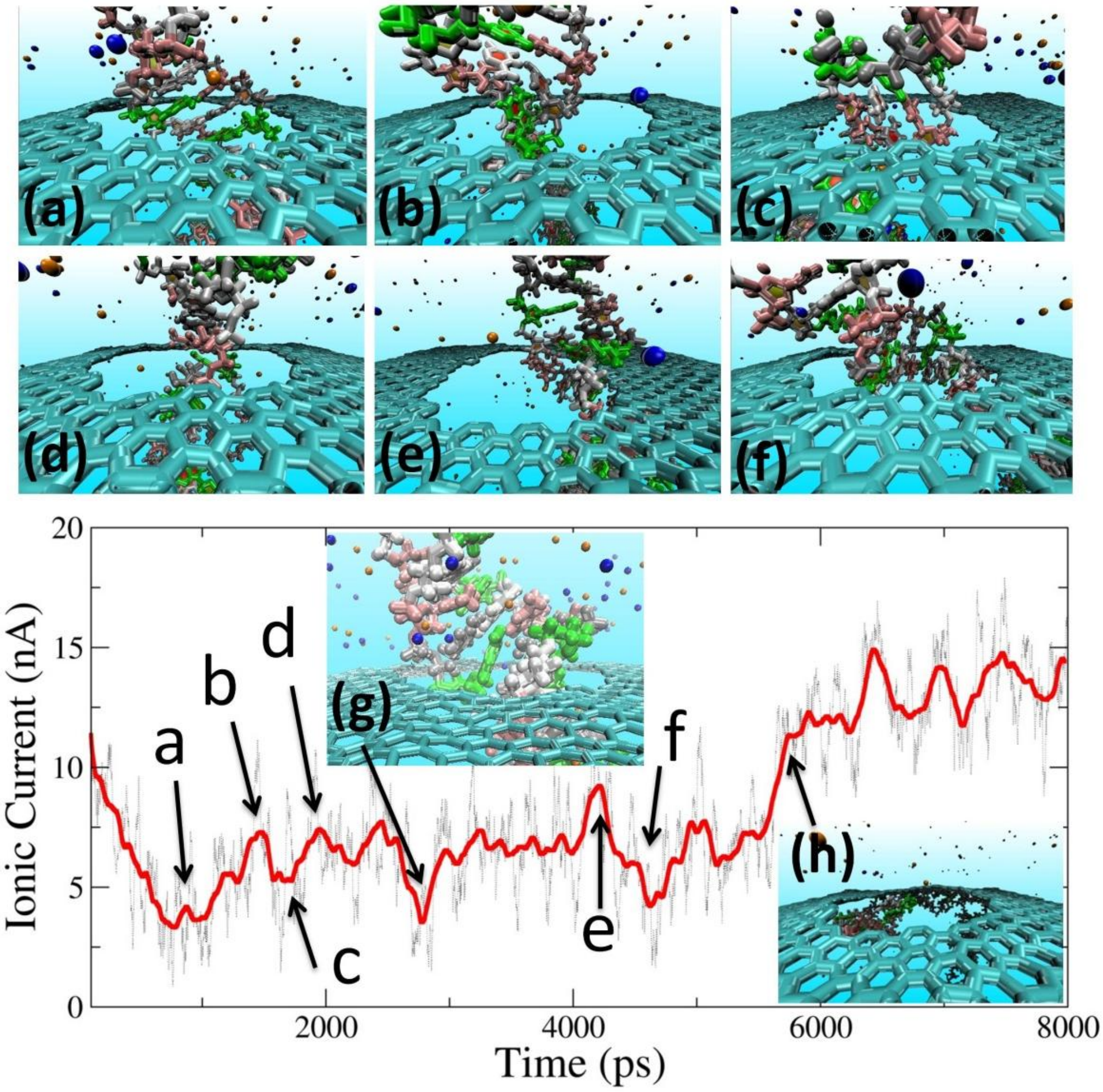

Figure 3. Both of the original (grey line) and smoothed (red line) ionic current were plotted as function of simulation time. The insets (a-h) show the local conformations of DNA in graphene nanopore, which is corresponding to the undulate of ionic current (marked with arrows a-h).

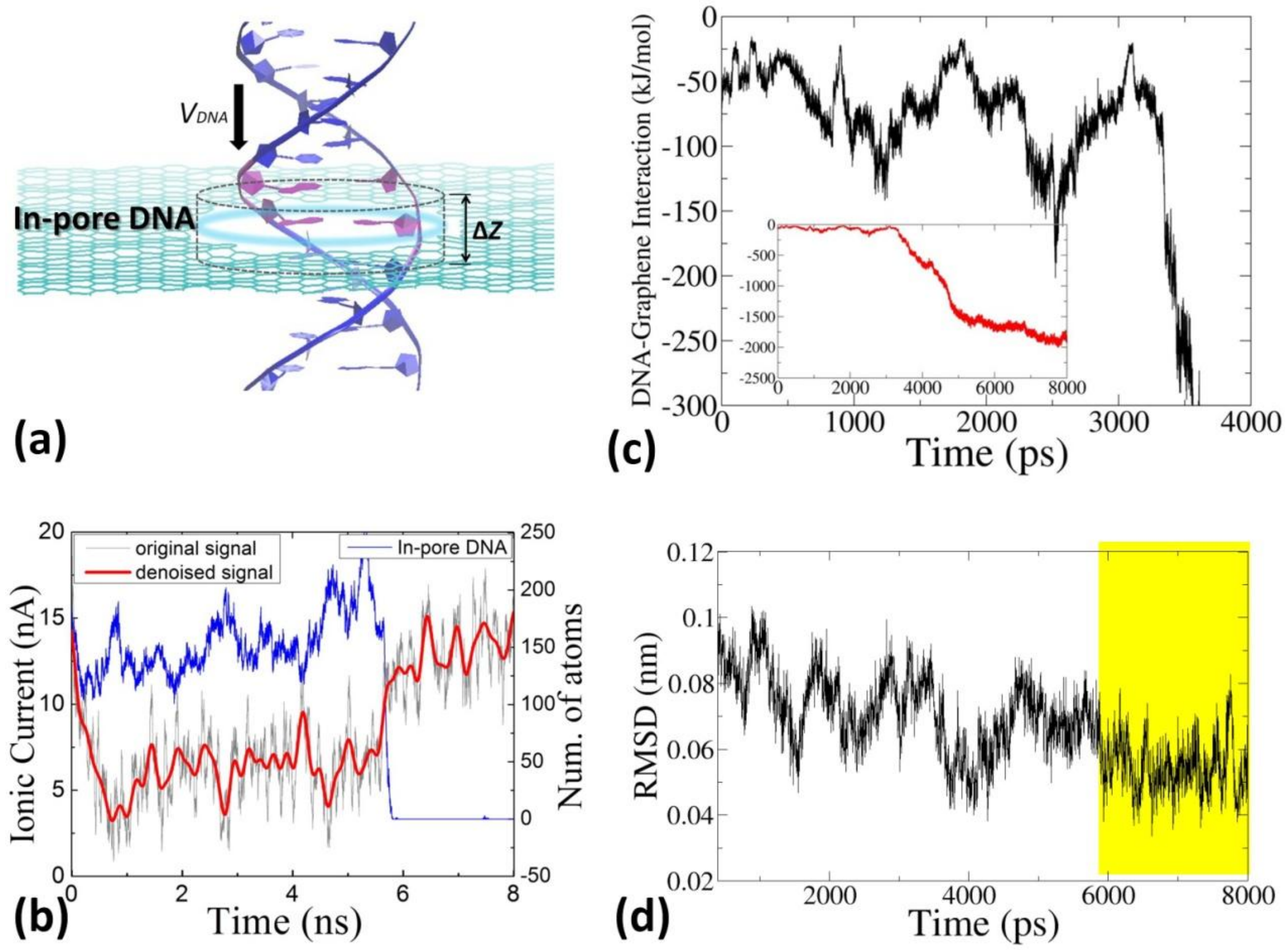

Figure 4. (a) The representation of in-pore DNA (colored in violet), $\Delta Z$ was set to 1 nm. (b) The time evolutions of the accumulated number of atoms of DNA accumulated in graphene nanopore (blue line) and ionic current signals (red line and grey line) of the simulation [2V, R1]. (c) The time evolution (0-4 ns) of the interaction energy between DNA and graphene of simulation [2V, R1]. The DNA-graphene interaction energy of whole trajectory was also showed as inset (0-8ns). (d) The time evolution of the root-mean-square distance (RMSD) of graphene nanopore of simulation [2V, R1]. The yellow band highlighted that the magnitude of RMSD of graphene nanopore was reduced after the translocation of DNA has finished (about 6-8ns).

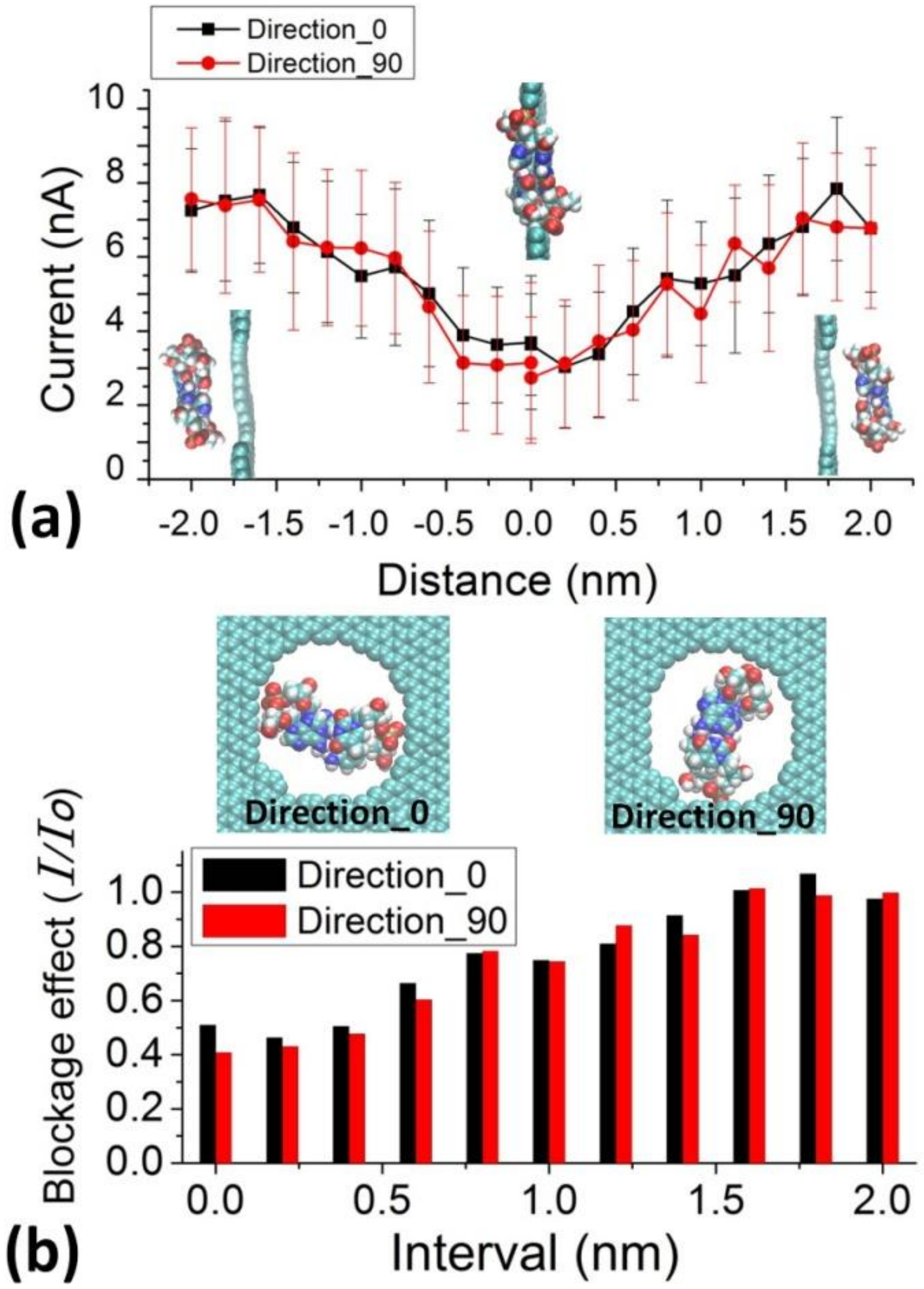

Figure 5. (a) The ionic current was plotted as function of the separation between DNA fragment and graphene nanopore. (b) The blockage effect of DNA to ionic current ($I/I_o$) with two orientations was plotted as function of interval between DNA and graphene nanopore. $I_o$ is the open pore ionic current of the graphene nanopore. The schematic diagrams of the position and orientation altering of DNA fragment in ionic current calculations were shown as insets.